\documentclass[]{article}
\usepackage{amsmath}
\usepackage[square,comma,numbers]{natbib}
\usepackage{graphicx}
\usepackage[scriptsize,bf]{caption}
\usepackage{authblk}

\newcommand{\br}[1]{\left(#1\right)}
\newcommand{\sq}[1]{\left[#1\right]}
\newcommand{\ud}[1]{\mathrm{d}{#1}}

\newcommand{\Arccos}[1]{\mathrm{Arccos}{\,#1}}
\newcommand{\Arcsin}[1]{\mathrm{Arcsin}{\,#1}}

\usepackage{amsmath}
\numberwithin{equation}{section}

\usepackage{color}

\usepackage[normalem]{ulem}
\usepackage{soul}
\setstcolor{red}

\definecolor{mygreen}{rgb}{0,0.4,0}

\begin{document}

\title{Nakedly singular counterpart of Schwarzschild's incompressible star. A barotropic continuity condition in the center.}

\author[1]{{\L}ukasz Bratek}
\author[1]{Joanna Ja{\l}ocha}
\author[1,2]{Andrzej Woszczyna}

\affil[1]{Institute of Physics, Cracow University of Technology, ul. Podchor\c{a}\.{z}ych 1, PL-30084 Krak\'{o}w, Poland}
\affil[2]{Copernicus Center for Interdisciplinary Studies,
ul. S{\l}awkowska 17, PL-31016 Krak\'{o}w, Poland}

\date{\today}

\maketitle

\begin{abstract}
A static sphere of incompressible fluid with uniform proper energy density is considered  as an example of exact star-like solution with  weakened central regularity conditions characteristic of a nakedly singular spherical vaccuum solution. The solution is a singular counterpart of the Schwarzschild's interior solution. The initial condition in the center for {general} barotropic equations of state is established.
\end{abstract}

\newcommand{\rad}{R_s}
\newcommand{\cmR}{R}
\newcommand{\cmW}{W}
\newcommand{\cmM}{M}
\newcommand{\singM}{{\cal M}_g}
\newcommand{\smM}{{\cal M}}

\section{Introduction}

Curvature singularities can be characterised in terms of scalars formed out of the Riemann tensor. A set of $16$ independent real scalars is provided by Carminati--McLenaghan invariants~\cite{bib:CM}. {An} arbitrary invariant of the curvature tensor can be expressed in terms of these scalars. For perfect fluid solutions at most $9$ of the scalars are independent, depending on the Petrov type. 

For static spherically symmetric {perfect fluid solutions of Einstein equations} \cite{bib:OV} one can verify by direct calculation that 
only $3$ of {the} {curvature} scalars are independent, {for example} $R^a{}_a$, $S^a{}_b\,S^b{}_a$ and $C^{ab}{}_{cd}\,C^{cd}{}_{ab}$ (here $R_{ab}$ is the Ricci tensor, $S_{ab}$ is the Pleba{\'n}ski tensor and $C_{abcd}$ is the Weyl tensor). Then all of {the} non-vanishing Carminati--McLenaghan invariants are functions of the three scalars.
 Up to constant factors, the three scalars are respectively:
$\br{\rho(r)-3\,p(r)}$, $(p(r)+\rho(r))^2$ and $r^{-6}\br{M(r)+M_g-\frac{4}{3}\pi\,r^3\rho(r)}^{{2}}$, {where $M(r)=4\pi\int\limits_0^r\tilde{r}^2\rho(\tilde{r})\ud{\tilde{r}}$ and $M_g$ is an integration constant with the dimension of mass.}
 The invariants $R^a{}_a$, $S^a{}_b\,S^b{}_a$ involve only {the} density $\rho(r)$ and pressure $p(r)$ {and therefore} are related to each other by the equation of state. 
In {the investigated context there are} thus only $2$ functionally independent curvature scalars, {one may} chose {$R^a{}_a$} and $C^{ab}{}_{cd}\,C^{cd}{}_{ab}$. 
The conformal invariant $C^{ab}{}_{cd}\,C^{cd}{}_{ab}$ {involves} the 
Misner--Sharp mass function \cite{bib:MS_1963} which is defined for {a} time-dependent spherically symmetric system as a purely geometric quantity $$
m(r,t)=\frac{\tilde{R}(r,t)}{2G}\br{1-g^{ab}\partial_{a}\tilde{R}(r,t)\partial_{b}\tilde{R}(r,t)}$$ with $\tilde{R}$ being the areal radius and $(t,r)$ the co-moving coordinates of the fluid. For the static solution of {interest in this paper one has} simply $\tilde{R}(r,t)=r$ and so $m(r,t)=M(r)+M_g$.  Then, the Misner--Sharp mass function splits into two parts: 
$M(r)$ {measuring} the amount of continuous matter enclosed within the areal radius $r$, and {$M_g$ being an independent constant parameter}. {For} a constant density solution, the conformal curvature invariant $C^{ab}{}_{cd}\,C^{cd}{}_{ab}$ {reduces to} $\frac{48\,G^2\,{M_g}^2}{c^4\,r^6}$, formally coinciding with the result for Schwarzschild's black hole. When $M_g\neq0$, this function of $r$ contributes as  the leading term {in} the conformal invariant, making it to blow up in the limit $r\to0$. In this sense, $M_g$ is the conformal curvature counterpart of mass independent of {the} matter content.

The regularity requirements for the Riemann and metric tensors \cite{bib:Wald} are {usually} imposed as supplementary conditions on {the}
solutions for spherical gaseous clouds.
In particular, the {integration constant}  $M_g$  is assumed to vanish for star-like solutions regular at the center, unlike for the vacuum solutions with {the} central singularity.
{However, despite the assumed regularity, some of the gravitational collapse solutions have been observed to asymptotically settle to nakedly singular equilibrium configurations}  \cite{bib:Joshi_2012book,bib:Joshi_2011a,bib:Joshi_2011b,
bib:Joshi_2011c,bib:Joshi_2012,bib:Joshi_2013a,
bib:Joshi_2013b}.
{To model similar final endstates, or other hypothetical nakedly singular objects as such,} the {regularity} conditions might be relaxed by allowing {for} central singularities with $M_g<0$. 
The degree of irregularity of the curvature invariants are in this case the same as for the Schwarzschild black hole. 

With the relaxed conditions, {one is} faced with new interesting possibilities that could be tested. Take for example numerical
integration of the equation of structure of a bounded spherical cloud with the neutron star polytrope equation of state. The resulting ADM mass turns out to lie in the range exceeding the Oppenheimer-Volkoff limit \cite{Woszczyna2015}.
In this context {it would be} instructive to consider a simpler {exactly integrable} problem. Related to this is also the nature of the initial condition in the center that should be imposed on solutions with general barotropic equation of state. {As shown in the next section,} the presence of the degree of freedom $M_g<0$ modifies the analytic properties of the Tolman-Oppenheimer-Volkoff
 equation \cite{bib:OV, tolman1939} to such {an} extent that the pressure required to keep in equilibrium a
positive density barotropic matter becomes negative for radii small enough, unless both the density and pressure vanish in the center. {As is found later,} with a given barotrope it may be impossible to obtain a solution satisfying these conditions (as it {occurs} for a polytrope $p=\kappa\, \rho^{\Gamma}$ with $\Gamma>1$). This imposes a limitation on the forms of barotropic matter with regular pressure profile at the center that could support stars against disintegration.

The pressure profile of a conformally singular incompressible spherical cloud of uniform proper density {is found in an analytical way in Sec.\ref{sec:2}}. For large radii the pressure behaves as expected for a star-like solution, decreasing to zero with growing distance from the center. However, as one goes in the opposite direction towards the singularity, the pressure, after attaining a maximum,  gradually decreases to zero and then becomes negative, attaining a unique value in the center. This solution is a {nakedly} singular counterpart of the non-singular Schwarzschild's interior solution \cite{bib:2016karl}.

\section{\label{sec:1} The central boundary condition}
The TOV triad of {the} equilibrium equations for a {conformally singular} spherical body consisting of baroropic material reads: 
\begin{eqnarray}\label{eq:tov} \chi '(x) &=& - \frac{\left( -\mu  + \mathcal{M}(x) + x^3\,\chi (x) \right) \,
       \left( \chi (x) + \omega[\chi (x)] \right) }{x\,
       \left( x + 2\,\mu  - 2\, \mathcal{M}(x) \right) },\quad \mu>0,\nonumber\\
 \mathcal{M}'(x) &=& x^2\,\omega[\chi(x)],\qquad  \mathcal{M}(0)=0,\\
f(\chi,\omega)&=&0.\nonumber\end{eqnarray}
The equations involve three dimensionless functions $\chi$, $\omega$ and $\mathcal{M}$  of a dimensionless variable $x=r/r_o$ ($r$ is the areal radius coordinate). The functions describe, respectively, the pressure profile: $p(r)=p_o\,\chi(r/r_o)$, the mass density profile: $\rho(r)=\rho_o\,\omega[\chi(r/r_o)]$, and the integrated mass profile: $\mathcal{M}(r/r_o)=\frac{4\pi}{m_o}\int_{0}^r \tilde{r}^2\rho(\tilde{r})\mathrm{d}{\tilde{r}}$, $\mathcal{M}(0)=0$. Functions $\omega$ and $\chi$ are not independent but related by some equation of state $f(\chi,\omega)=0$, that is, $f(\chi(x),\omega[\chi(x)])\equiv0$ for any $x$.  The dimensional parameters $p_o$, $\rho_o$, $m_o$ and $r_o$ are set so as to eliminate all inessential constants from the equations.\footnote{In writing eqs.\ref{eq:tov} we have assumed: $\frac{4\pi G r_o^2 p_o}{c^4} =1$, $\frac{\rho_o c^2}{p_o}=1$ and $\frac{G m_o}{r_o c^2}=1$.} The only free parameter left, the $\mu$, measures in units of $m_o$ the negative mass {parameter in the conformal curvature}, hence the total Misner-Sharp mass function is $m_o\br{\mathcal{M}(r/r_o)-\mu}$ ({it is assumed that} $\mu>0$).

The standard form of the  metric tensor of a static and spherically symmetric spacetime in which the equilibrium equation acquires the above form, implies that the term $x + 2\,\mu  - 2\, \mathcal{M}(x)$ in the denominator of the expression for $\chi'(x)$ should be everywhere positive for a star-like solution.
In the neighbourhood of $x=0$, this will be possible for a bounded $\omega$ only with $\mu\geq0$. On the contrary, with $\mu<0$, either the condition that the velocity four-vector of the fluid be everywhere time-like could not be satisfied, or the signature of the metric form would be inappropriate for a genuine spacetime, even though Eqs.\ref{eq:tov} could be formally the same. Moreover, the asymptotic flatness with positive total mass 
requires that $\mathcal{M}(\infty)-\mu>0$.
As follows from the expression for $\chi'(x)$  in Eqs.\ref{eq:tov}, the pressure will be a decreasing function of $x$, like for an ordinary star in {the} equilibrium, only for $x$ large enough. For lower $x$,  $\chi'(x)$ may become positive and the pressure attain a local maximum. Such a maximum {has been} observed for {a} numerical model studied in \citep{Woszczyna2015}.

In finding solutions, especially numerical ones in the $x=0$ vicinity, it may prove helpful the following 
\begin{description}\item[Theorem.]
{\it If the pressure $\chi(x)$ is continuous at $x=0$ and continuously differentiable for $x>0$ (as expected for a star consisting of ordinary matter) then $\lim_{x\to0}\br{\chi(x)+\omega[\chi(x)]}=0$ {for $\mu>0$} (implying that either $\chi(0)=0=\omega(0)$, or $\chi(0)=-\omega(0)$ for $\omega(0)\ne0$).}
\end{description}
  {Proof. Making use of the mean value theorem applied to function $\chi(x)$}, {it follows} with the use of Eqs.\ref{eq:tov} that 
$$\chi(x)-\chi(0)=\frac{\chi(x\,\eta_x)+\omega[\chi(x\,\eta_x)]}{\eta_x}\cdot\frac{\mu-\mathcal{M}(x\,\eta_x)-\eta_x^3\,x^3\chi(x\,\eta_x)}{2\,\mu+x\,\eta_x-2\mathcal{M}(x\,\eta_x)},$$
where $\eta_x$ is an $x$-dependent number such that $0<\eta_x<1$ and $\eta_x\to0$ as $x\to0$. With bounded $\chi$ and $\omega$, both $\mathcal{M}(x)$ and $x^3\chi(x)$ tend to $0$. As so, these functions can be made arbitrarily small numbers compared with $\mu$ for any $x$ small enough. Now, taking the limit in the above formula, {one obtains}
$$0=\lim\limits_{x\to0}\frac{\chi(x\eta_x)+\omega[\chi(x\eta_x)]}{2\,\eta_x}
\quad\Rightarrow\quad \chi(0)+\omega[\chi(0)]=0,$$ which ends the proof. With $\mu=0$ the above implication would not follow, because then $\lim\limits_{x\to0}\frac{0-\mathcal{M}(x)-x^3\chi(x)}{0+x-2\mathcal{M}(x)}=\lim\limits_{x\to0}\frac{-\mathcal{M}'(x)-3x^2\chi(x)-x^3\chi'(x)}{1-2\mathcal{M}'(x)}=0$ if only $x^3\chi'(x)\to0$, with the same assumptions as before.

\medskip

In accordance with the theorem, for barotropes with non-negative $\omega$ and with $\chi$ continuous at $x=0$, {one infers} that $\chi(0)=-\omega[\chi(0)]<0$ or $\chi(0)=0=\omega[0]$. For instance, the pressure becomes negative in a neighbourhood of the center for the exact solution presented in Sec.\ref{sec:2}. 
But 
negative pressure is impossible for matter described by a polytropic equation of state for which both the density and pressure are required non-negative. The only 
possibility for such a polytrope is that either $\chi(0)=0$ as follows from the above theorem, or the reservations of the theorem are not met for a given solution. On that account, assume that in the limit of low pressure, matter can be described by a polytrope $\omega(\chi)=a\chi^\gamma$ with $\gamma>0$ ({usually it is expected that} $\omega(\chi)\sim a\,\chi^{n/(n+1)}$, $n>0$ for realistic polytropes). With $0<\gamma<1$, like for ordinary matter, $\chi+a\,\chi^\gamma\sim a\,\chi^\gamma$ as $\chi\to0$, then, upon integration of the approximated equation 
$\chi'(x)\approx(\chi(x)+\omega[\chi(x)])/(2x)$, {one gets}  $\chi^{1-\gamma}\sim a(1-\gamma)\ln(\sqrt{x})+c$ -- a contradiction with the assumed initial condition $\chi(0)=0$ (again, in writing the approximated equation {one makes use of} the fact that $\mathcal{M}$ and $x^3\chi$ are negligible compared to $\mu$ for finite $\chi$ and $x$ small enough). Hence, a solution with a polytropic equation of state  and finite pressure would be possible at $x=0$ only for $\gamma\geq1$, in which case  $\chi\sim c \sqrt{x}$ as $x\to0$. For a polytrope with $0<\gamma<1$ a solution with vanishing initial pressure would have to start at some $x=x_o>0$ and behave like $\chi\sim c (x-x_o)^\frac{1}{1-\gamma}$ as $x\to x_o^+$. A similar line of reasoning for the particular case of polytropic equations of state, showing that $\chi(0)=0$ for $\gamma\geq 1$ and that finite positive $\chi(0)$ is incompatible with $0<\gamma<1$,  was presented by Oppenheimer and Volkoff \cite{bib:OV}.
The theorem {as given above} for barotropic equations of state is more general.

\section{\label{sec:2}Interior solutions }

{This section considers static equilibrium of a conformally singular in the centre} spherical star consisting of incompressible fluid {with} constant proper mass density. 
Let $a$ be the areal radius of the star, $M$ the total mass  of the regular mass distribution of the star, and $-M_o$ ($M_o>0$) the negative mass {parameter in the conformal curvature invariant}. The three quantities determine two independent dimensionless parameters  $\alpha$ and $\mu$ of the model 
$$\alpha:=\frac{a\,c^2}{2\,G\,M}>1-\mu,\qquad \mu:=\frac{M_o}{M}\in(0,1)$$
(the condition $\alpha>1-\mu$ will become clear later).
In this model the dimensionless pressure profile $\chi$ and the uniform density profile $\omega$ are defined by  
$$p(r)=p_o\cdot \chi(r/a)\quad \mathrm{and}\quad \frac{\rho(r)}{\rho_o}=\omega(\chi(x))\equiv1, \qquad \mathrm{with}\quad \frac{p_o}{c^2}=\rho_o=\frac{M}{\frac{4}{3}\pi a^3}.$$
{The areal radius $a$ can be chosen as the unit of length:} $r_o=a$. Accordingly, $x=r/a$ is the dimensionless radial variable and the interval $0<x<1$ describes the star interior.  The  {Misner-Sharp} mass function is $\mathcal{M}(x)-\mu=x^3-\mu$ for $0<x<1$ {(it can be verified that Komar mass will be different and involve also $\chi(x)$, however, both masses will overlap at $x=1$ and be equal to $1-\mu$)}. In these units, {for the general form of the line element
$$\ud{s}^2=-e^{2\,\Psi (x)}\,c^2{\ud{t}}^2 +a^2\br{ 
  \frac{ {\ud{x}}^2 }{1 -\frac{\mathcal{M}(x)-\mu}{\alpha\,x}}
  + x^2\,\left( {\ud{\theta}}^2 + 
     {\sin^2{\theta}}\,{\ud{\phi}}^2 \right)},$$ }the equilibrium equation to be satisfied by the pressure profile $\chi(x)$ acquires the following form
\begin{equation}\label{eq:singschw}\chi '(x) = -\frac{ \left( 1 + \chi (x) \right) \,
       \left( x^3\left( 1 + 3\,\chi (x) \right)-\mu  \right) 
        }{2\,x\,\left( -x^3 + x\,\alpha  + \mu  \right) }, \qquad 
        \left.\begin{array}{c}0<x<1,\\ \chi(1)=0.\end{array}\right.\end{equation}
  {Once the equation is solved, the other metric function satisfying the equation $\Psi '(x) = - \frac{\chi '(x)}{1 + \chi (x)}$ is easily found
$$e^{2\,\Psi (x)}=\frac{C}{\br{1+\chi(x)}^2}, \qquad C=1+\frac{\mu-1}{\alpha},$$ where the integration constant $C$ is set by the requirement that} {f}or $x>1$ the resulting metric must be extended by the external Schwarzschild's vacuum metric with total mass $1-\mu>0$. {The conformal curvature scalar for this solution is
$$C^{ab}{}_{cd}\,C^{cd}{}_{ab} =\frac{48\, G^2 M^2}{c^4\,r^6}\cdot\left\{
\begin{array}{ccc}
\mu^2&,& 0<r<a\\
\br{1-\mu }^2&,& r>a
\end{array}\right.$$ and diverges at the center. The invariant  is discontinuous on the star boundary $r=a$ unless $\mu=1/2$. }

\subsection{Nonsingular interior solution ($\mu=0$)}
The solution for $\mu=0$ was found by  Schwarzschild in 1916 \citep{bib:2016karl}. It is known as Schwarzschild's interior solution. In our notation the solution reads:
\begin{equation}\chi(x)|_{{\mu=0}}=\frac{\sqrt{\alpha-x^2}-\sqrt{\alpha-1}}{3\sqrt{\alpha-1}-\sqrt{\alpha-x^2}}, \qquad \alpha>\frac{9}{8}.\label{eq.intschw}
\end{equation}
The above condition for $\alpha$ assures that the pressure is finite (and everywhere non-negative). Although there is a region of negative pressure for $\alpha<9/8$, such solutions are of interest today. As noticed 
by Mazur and Mottola \cite{Mazur2015}, the divergence in pressure at $x=3\sqrt{1-\frac{8}{9}\alpha}$ is integrable and the solution with {the} distinguished value $\alpha=1$  can be reinterpreted as a  gravitational condensate star with constant negative pressure $\chi=-\omega=-1$ throughout the star and some residual transverse stresses at the boundary of the star.    

\subsection{{Nakedly} singular interior solution ($\mu>0$)}
The Schwarzschild's interior solution is qualitatively changed in the $x=0$ vicinity if $\mu>0$.  {Then}, for $\chi$ continuous at $x=0$, there is a $\delta>0$ such that   $\chi'(x)\sim(1+\chi(x))/(2x)$ for $x<\delta$, hence
$\chi(x)\sim -1+A\sqrt{x}$, proving that $\chi(0)=-1$. 
This is consistent with the theorem given in Sec.\ref{sec:1}. Note that $\chi(0)=-1$, independently of the integration constant $A$ the value of which is controlled by the other part of the solution for larger $x$. Namely, $A$ in the asymptotic form of $\chi$ at $x=0$ is determined by the value $\chi=0$ at the boundary $x=1$. 

A substitution  $$ \chi(x)=\frac{2}{3\,u(x)}-1 $$ 
      transforms the non-linear equation Eq.\ref{eq:singschw} for $\chi$, to a linear equation for $u$, with rational functions  as coefficients:
\begin{equation}\label{eq:linear} u'(x)+\frac{2\,x^3 + \mu}
     {2x\left( -x^3 + x\,\alpha  + \mu  \right) }\,u(x)  = 
   \frac{x^2}{-x^3 + x\,\alpha  + \mu }.\end{equation}
   
Before proceeding further, the roots of the denominators in the linear equation {have to be examined}.
In this respect it is convenient to introduce the following combination of numbers $\alpha$ and $\mu$: 
$$\upsilon=\sqrt{\frac{27\mu^2}{4\,\alpha^3}},\qquad \mathrm{or\ equivalently} \quad \upsilon=
\frac{M_o}{2M}\br{\frac{6\, G M}{a\,c^2}}^{3/2}.$$
{A} new dimensionless parameter $\upsilon$ can be regarded as a discriminant of the trinomial $w(x)=-x^3+\alpha\,x+\mu$ characteristic of {the investigated} problem.    For $\upsilon<1$ the roots of the qubic equation $w(x)=0$ are all real and read
     $${x_k}=2\sqrt{\frac{\alpha}{3}}\cos\br{\frac{2k\pi}{3}+\frac{1}{3}\,\Arccos{\upsilon}},\quad k=0,1,2,\quad \upsilon<1.$$ 
Then $x_0>\sqrt{\alpha}$,   $x_2<0$ and $x_3<0$. For 
     $\upsilon>1$ only one root is real and reads $2\sqrt{\frac{\alpha}{3}}\cosh\br{\frac{1}{3}\,\mathrm{Arcosh}{\,\upsilon}}>\sqrt{\alpha}.$     
           For physical reasons, it is important that the roots be located exterior to the stellar interior $0<x\leq1$. This will be the case if {one safely assumes} independently of $\mu$ that $\alpha>1$, that is, when 
     the coordinate radius of the star is greater than its Schwarzschild's radius: $a>\frac{2\,G\,M}{c^2}$. In fact, with a given $\mu$ the $\alpha$ could be assumed lower: $\alpha>1-\mu$, then
     $w(x)>-x^3+(1-\mu)x+\mu>0$ for any $0<x<1$ and $0<\mu<1$.
      
\begin{figure}
       \centering
        \includegraphics[width=0.8\textwidth]{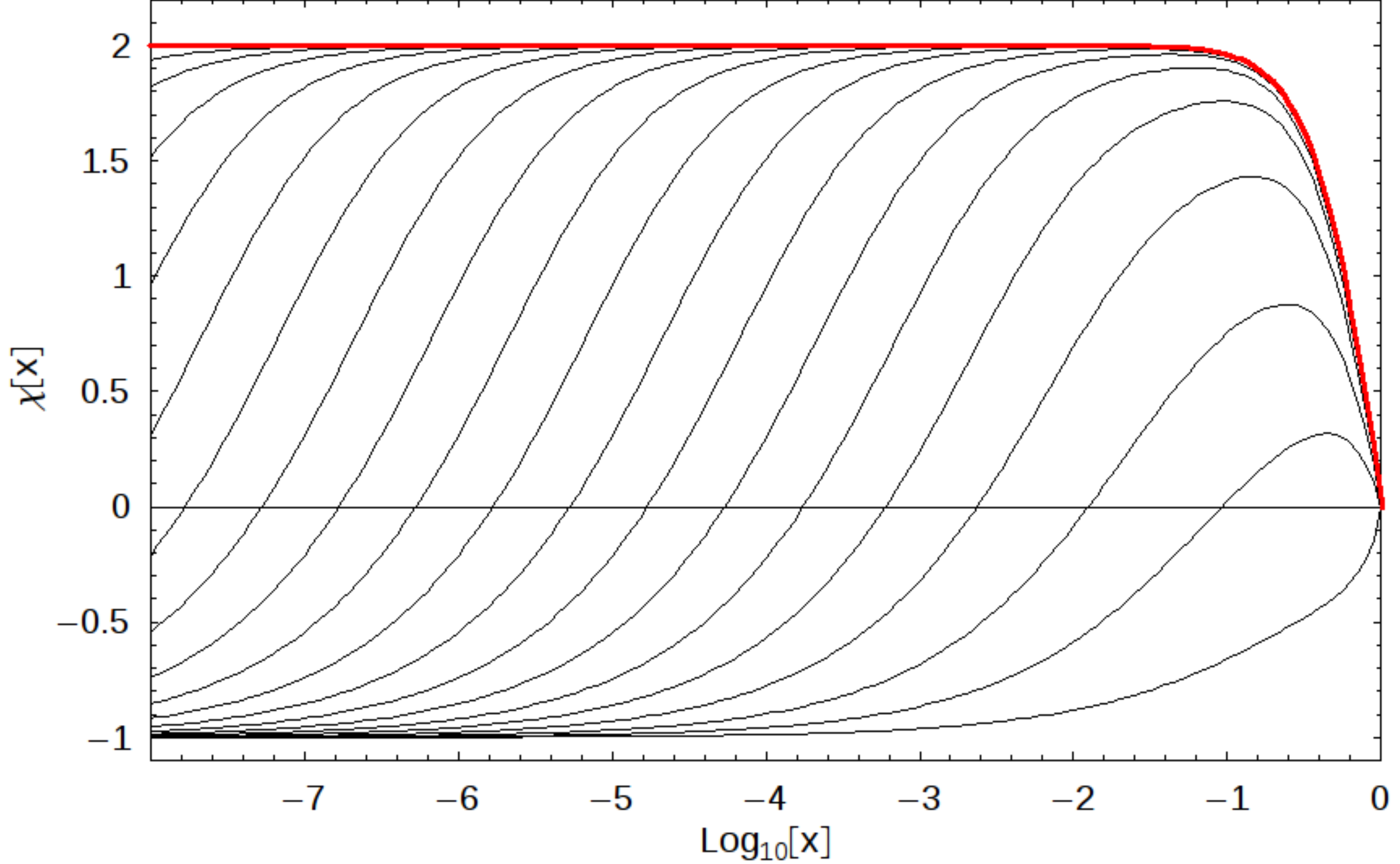}
  \caption{\label{fig.SingularUniform}The exact pressure profile $\chi(x)$ for the constant density ball with central {conformal} 
  singularity, shown for various 
     $\mu>0$ [{\it black lines}]. For $x>0$, as $\mu\to0$, the profiles tend (non-uniformly) to the $\mu=0$ regular Schwarzschild's {interior solution} profile [{\it red line}], while $\chi(0)$ remains fixed by $-1$ for any $\mu>0$.  }
       \end{figure}

Coming back to  equation \ref{eq:linear}, with the reservations for the roots made, {the linear equation may be recast} into the equivalent form
   $$\frac{\partial }{\partial x}\frac{\sqrt{x}\, u(x)}{\sqrt{-x^3+x\,\alpha +\mu }}=\frac{x^{5/2}}{\left(-x^3+x\,\alpha +\mu \right)^{3/2}},$$ {hence} the general solution involves an elliptic integral and reads
     $$u(x) = {\sqrt{\frac{-x^3 + x\,\alpha  + \mu }{x}}}
   \int \frac{x^{\frac{5}{2}}}
       {{\left( -x^3 + \alpha\,x  + \mu  \right) }^{\frac{3}{2}}}\,dx.$$
        In terms of  a new integration variable
  \begin{equation}\label{eq:sx}s(x)=-\frac{1}{2}-\frac{\upsilon}{x}\,\sqrt{\frac{\alpha}{3}},\end{equation} the solution can be expressed in a form with standardized elliptic integral          
$$\frac{u(x)}{\sqrt{\frac{-x^3 + x\,\alpha  + \mu }{\alpha\,x}}}=C+\frac{{\upsilon }^2\sqrt{6}}{4}\,\mathcal{I}(x),\qquad \mathcal{I}(x)=
      \int\limits_{-\infty}^{s(x)}\frac{\mathrm{d}\tilde{s}}{\sqrt{(s_3-\tilde{s})^3(s_2-\tilde{s})^3(s_1-\tilde{s})^3}}.$$
Parameters $s_n$ in the integrand 
are given by      
\begin{equation}\label{eq:sns}s_n=\cos\br{\frac{2}{3}\Arcsin{\upsilon}+\frac{2n\pi}{3}},\qquad n=1,2,3.\end{equation}
They are the roots of the qubic form $-s^3+\frac{3}{4}\,s+\frac{1-2\,\upsilon^2}{4}\equiv(s_3-s)(s_2-s)(s_1-s)$,  ($s_3>s_2>s_1$).         
  {They} correspond to the roots of the previous qubic form  $w(x)$: $x_k=-\frac{2\,\upsilon}{1+2\,s_{k+1}}\sqrt{\frac{\alpha}{3}}$, $k=0,1,2$.
The integration constant $C$
  is set by the condition $\chi(1)=0$ ($u(1)=2/3$) on the surface of the star. Hence,
the final solution for $0<x<1$  reads
\begin{equation}\label{eq:final}\chi (x) = \frac{{\sqrt{\frac{x\,\left( -1 + \alpha  + \mu  \right) }
         {-x^3 + x\,\alpha  + \mu }}}}{1 - 
      \frac{81\,{\sqrt{6}}}{32}\,\frac{\mu^2}{\alpha^3}\,
       {\sqrt{\frac{-1 + \alpha  + \mu }{\alpha }}}\,\left( \mathcal{I}(1) - \mathcal{I}(x) \right) } - 1, \qquad \alpha>1-\mu.\end{equation}
       The explicit form of the solution  is very intricate; for clarity of the presentation the integral $\mathcal{I}(x)$ is given in its entirety in Appendix \ref{app:1}. To get some idea of how the pressure profile looks like anyway, {it is presented} in the graphical form in   Fig.\ref{fig.SingularUniform}. Outside the center $x>0$, the pressure tends point-wise in the limit $\mu\to0$ to the interior Schwarzschild's solution Eq.\ref{eq.intschw} with $\mu=0$. Since for $\mu>0$
       the pressure at $x=0$ is fixed by a negative number $\chi(0)=-1$ independent of $\mu$, this limit is non-uniform.

\appendix
\section{\label{app:1}Appendix}
Let $s_1$, $s_2$ and $s_3$ be defined as in  Eq.\ref{eq:sns}, the function $s(x)$ as in Eq.\ref{eq:sx}.
The definite improper integral $\mathcal{I}(x)$  present in the expression for the  pressure profile in Eq.\ref{eq:final}, has standard form of an elliptic integral that can be found in tables of integrals, and it reads (see for example integral 3.136.1 page 263 in \cite{bib:tables}):
\begin{equation*}
\begin{aligned}
\mathcal{I}(x):=&\int\limits_{-\infty}^{s(x)}\frac{\mathrm{d}\tilde{s}}{\sqrt{({s_3}-\tilde{s})^3({s_2}-\tilde{s})^3({s_1}-\tilde{s})^3}}=\\
&\frac{2}{({s_3}-{s_2})^2({s_2}-{s_1})^2\sqrt{({s_3}-{s_1})^3}}
\times\Big\{({s_2}-{s_1})({s_3}+{s_2}-2{s_1})F(\varphi(x),\kappa)\\
&\phantom{\hspace{0.2\linewidth}}-2\br{{s_1}^2+{s_3}^2+{s_2}^2-{s_3}{s_2}-{s_3}{s_1}-{s_2}{s_1}}E(\varphi(x),\kappa)\Big\}\\
&+\frac{2\sq{{s_1}({s_3}-{s_1})+{s_2}({s_3}-{s_2})-s(x)\br{2{s_3}-{s_1}-{s_2}}}}{({s_3}-{s_2})({s_3}-{s_1})({s_2}-{s_1})^2\sqrt{({s_3}-s(x))({s_2}-s(x))({s_1}-s(x))}},
\end{aligned}
\end{equation*}
where 
$$
\varphi(x)=\Arcsin{\sqrt{\frac{{s_3}-{s_1}}{{s_3}-s(x)}}}, \qquad \kappa=\sqrt{\frac{{s_3}-{s_2}}{{s_3}-{s_1}}}.   
$$
Here, $F$ and $E$ are the incomplete elliptic integrals of the first and second kind, as defined in the same tables of integrals \cite{bib:tables}:
$$F(\phi,k)=\int\limits_0^{\phi}\frac{\mathrm{d}{\theta}}{\sqrt{1-k^2\sin^2\theta}}, \qquad 
E(\phi,k)=\int\limits_0^{\phi}\sqrt{1-k^2\sin^2\theta}\,\mathrm{d}{\theta}.$$
The form of the expression on the right hand side of the equality sign in the above definition of $\mathcal{I}(x)$ assumes that  ${s_3}>{s_2}>{s_1}>s(x)$.

\bibliography{singular_interior}
\bibliographystyle{unsrt}

\end{document}